\documentclass[10pt,aps,prl,twocolumn,showpacs,amsmath,superscriptaddress]{revtex4-1}

\usepackage{amsmath}
\usepackage{graphicx}% Include figure files
\usepackage{dcolumn}% Align table columns on decimal point
\usepackage{bm}% bold math
\usepackage[colorlinks,linkcolor=blue,citecolor=blue,urlcolor=blue]{hyperref}

\begin{document}

% Title and authors

\title{Modulating carrier and sideband coupling strengths in a standing wave gate beam}

\author{Thomas E. deLaubenfels}
\affiliation{Georgia Tech Research Institute, Atlanta, GA 30332, USA}
\affiliation{School of Physics, Georgia Institute of Technology, Atlanta, GA 30332, USA}
\author{Karl A. Burkhardt}
\altaffiliation[Current address: ]{Department of Physics, University of Texas at Austin, Austin, TX 78712, USA}
\affiliation{School of Chemistry and Biochemistry, Georgia Institute of Technology, Atlanta, GA 30332, USA}
\author{Grahame Vittorini}
\altaffiliation[Current address: ]{Honeywell International, Golden Valley, MN 55422, USA}
\affiliation{School of Physics, Georgia Institute of Technology, Atlanta, GA 30332, USA}
%\author{Kenton R. Brown} 
%\affiliation{Georgia Tech Research Institute, Atlanta, GA 30332, USA}
%\affiliation{School of Physics, Georgia Institute of Technology, Atlanta, GA 30332, USA}
\author{J. True Merrill} 
\affiliation{Georgia Tech Research Institute, Atlanta, GA 30332, USA}
%\affiliation{School of Physics, Georgia Institute of Technology, Atlanta, GA 30332, USA}
%\author{Curtis Volin} 
%\affiliation{Georgia Tech Research Institute, Atlanta, GA 30332, USA}
%\author{Alexa W. Harter} 
%\affiliation{Georgia Tech Research Institute, Atlanta, GA 30332, USA}
\author{Kenneth R. Brown} 
\affiliation{School of Chemistry and Biochemistry, Georgia Institute of Technology, Atlanta, GA 30332, USA}
\affiliation{School of Physics, Georgia Institute of Technology, Atlanta, GA 30332, USA}
\affiliation{School of Computational Science and Engineering, Georgia Institute of Technology, Atlanta, GA 30332, USA}
\author{Jason M. Amini}
\affiliation{Georgia Tech Research Institute, Atlanta, GA 30332, USA}
%\affiliation{School of Physics, Georgia Institute of Technology, Atlanta, GA 30332, USA}

\date{\today}

% Abstract

\begin{abstract}
%The motion of trapped ions leads to sidebands in the ion's excitation spectra. The relative coupling strengths of the carrier and secular sideband transitions are usually fixed by the Lamb-Dicke factor and the ion temperature. 
We control the relative coupling strength of carrier and first order motional sideband interactions of a trapped ion by placing it in a resonant optical standing wave.  Our configuration uses the surface of a microfabricated chip trap as a mirror, avoiding technical challenges of in-vacuum optical cavities. %This is achieved by exciting the transition with a standing wave beam instead of the typical traveling wave. In such a configuration, the coupling strengths acquire an additional dependence on the ion's position within the standing wave.
Displacing the ion along the standing wave, we show a periodic suppression of the carrier and sideband transitions with the cycles for the two cases $180^\circ$ out of phase with each other.
% We suppress resonant carrier and sideband interactions by an amount equivalent to a $8$~dB and $11$~dB reduction in the beam power respectively.  
This technique allows for suppression of off-resonant carrier excitations when addressing the motional sidebands, and has applications in quantum simulation and quantum control.  Using the standing wave fringes, we measure the relative ion height as a function of applied electric field, allowing for a precise measurement of ion displacement and, combined with measured micromotion amplitudes, a validation of trap numerical models.

%This technique could be used to suppress off-resonant carrier excitations in two qubit gates, and the position dependence provides a measure of the ion displacement that can be used to map out the trapping potentials.
\end{abstract}

% 03.67.Lx -- Quantum computation
% 32.80.Qk -- Coherent control of atomic interactions with photons
% 37.10.Ty -- Ion traps

  % FIX THESE KEYWORDS WITH SOMETHING BETTER 6/2/2015
\maketitle

\begin{figure}[ht]
    \begin{center}
        \includegraphics{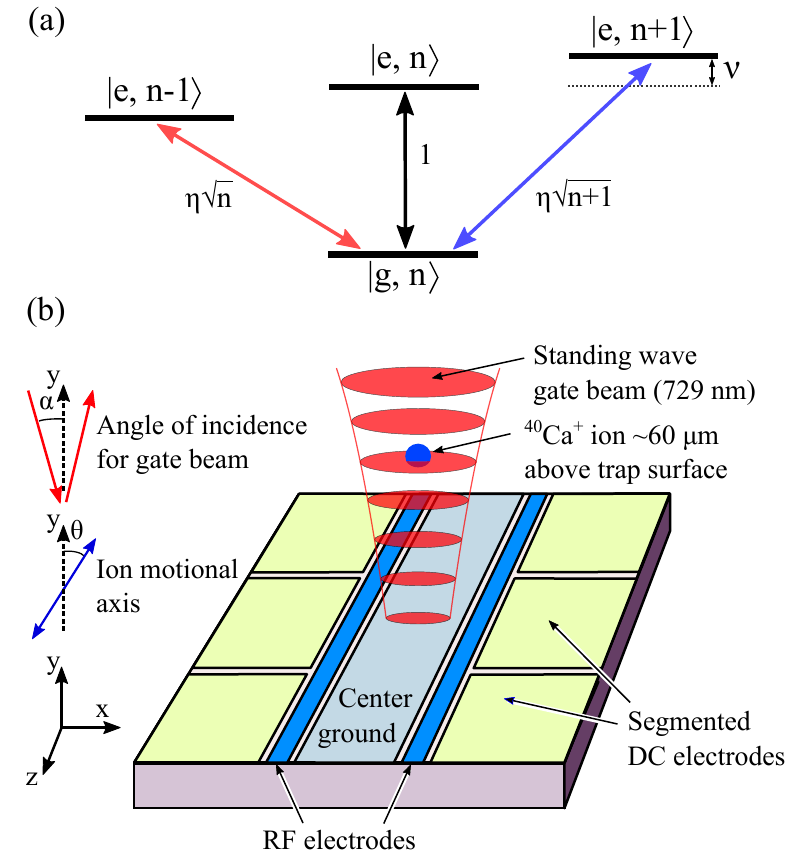}
		\end{center}
     \caption{\label{fig:schematic}(a) Energy diagram of an ion in a harmonic potential of frequency $\nu$, with arbitrary electronic states $|g\rangle$ and $|e\rangle$ and motional states $|n\rangle$. Carrier ($\Delta n=0$) and first order ($\Delta n=\pm 1$) sideband transitions shown. In a running wave configuration, the coupling strengths of the red and blue sidebands relative to the carrier are (to first order) proportional to $\eta \sqrt{n}$ and $\eta \sqrt{n+1}$, respectively, where $\eta$ is the Lamb-Dicke parameter.  (b) Schematic of our experimental configuration. $^{40}$Ca$^+$ ions are confined $\sim$60 $\mu$m above the trap surface by RF and DC electric fields. 729 nm light is used to excite the $|S_{1/2}, m_F = -\frac{1}{2} \rangle \rightarrow |D_{5/2}, m_F = -\frac{5}{2} \rangle$ quadrupole transition. The incident 729 nm laser (with incidence angle $\alpha$) is retroreflected from the aluminum trap surface to produce a standing wave. The ion is displaced through the standing wave fringes by changing the trap fields. In this configuration, the resulting carrier and sideband coupling strengths are represented by equations (1) and (2).}
\end{figure}

The excitation spectrum of a trapped ion in a radio frequency (RF) trap acquires sidebands due to the harmonic motion of the ion (Fig.~\ref{fig:schematic}(a))  \cite{Leibrandt09.PRL.103.103001}. The interaction between an optical field and the the trapped ion leads to an almost ideal Jaynes-Cummings interaction \cite{Jaynes63.IEEE.51.89,Wu97.PRL.78.3086} which couples the internal degrees of freedom to the ion motion and is the basis for two-qubit gates in ion trap quantum computation \cite{Cirac95.PRL.74.4091,Sorensen00.PRA.62.022311,Schmidt03.Nature.422.408}. Sideband interactions are used in trapped ion experiments for a variety of additional functions such as cooling to the motional ground state \cite{Diedrich89.PRL.62.403}, measurements of the ion heating rate \cite{Turchette.PRA.61.063418,Shu14.PRA.89.062308}, and identifying and cooling molecular ions \cite{Goeders13.JPhysChemA.117.9725,Rugango15.NJP.17.035009,Wan15.PRA.91.043425}. Off-resonant coupling to the carrier transition, either evident as motion independent population transfer or an AC Stark shift, places a limit on the speed of the sideband interactions.  Suppressing the carrier can remove this limit and, in particular, would allow for improved two-qubit gate fidelities as the gate time becomes comparable or shorter than a cycle of the harmonic motion \cite{Sorensen00.PRA.62.022311,Mizrahi14.APB.114.45}. 

Suppression of the carrier also has applications in quantum simulation.  Trapped ions have been proposed as a system for modeling the expansion of the universe \cite{Menicucci10.NJP.12.095019}. The simulation requires off-resonant excitation of both the red and blue sidebands by a red-detuned exciting field, with no coupling to the carrier. Because the blue sideband is both weaker and further from resonance than the carrier transition, suppression of the carrier is important for such an experiment.

Replacing running wave optical beams with standing wave beams provides a method to selectively suppress the carrier and reduce off-resonant excitations when addressing the motional sidebands \cite{Cirac92.PRA.46.2668,Zhang12.PRA.85.053420}. In such a configuration, the coupling strengths of the carrier and sidebands acquire a periodic dependence on the atom's spatial position within the standing wave fringes \cite{Cirac92.PRA.46.2668,James98.APB.66.181}, with the cycles for the two cases $180^\circ$ out of phase with each other. Standing wave beams have also been proposed for use in measuring parity nonconservation effects in trapped ions for this reason \cite{PhysRevLett.70.2383,0953-4075-36-3-320}. This periodic dependence of the coupling strengths has been demonstrated in cavity experiments with trapped ions \cite{Leibrandt09.PRL.103.103001,Guthohrlein01.Nature.414.49,Steiner13.PRL.110.043003,PhysRevLett.89.103001}. However, the use of cavities involves technical challenges, in particular when integrating the cavity with microfabricated ion traps where optics and dielectric mirrors can become charged \cite{Harlander10.NJP.9.093035,Clark14.PRAppl.1.024004}. 

In this paper, we demonstrate the same position dependence with a single mirror which, in this case, is simply the surface of the ion trap itself (Fig.~\ref{fig:schematic}(b)).  To handle the case of imperfect beam alignment, reflection losses, and similar system limitations, we extend the calculations of Refs.~\cite{Cirac92.PRA.46.2668,James98.APB.66.181} to the case of non-normal incidence laser beams and unequal couplings of the incident and reflected laser beams with the ion. We find a criterion for the out of phase carrier and sideband coupling strengths that is set by the incident angle of the laser beam and the orientation of the ion's harmonic motion. 

% \cite{Cirac95.PRL.74.4091,Wineland98.Fortsch.46.363,Haeffner08.PhysRep.469.155}

%The positional dependence of transition coupling strengths in standing wave fields is well documented in cavity experiments with trapped ions \cite{Leibrandt09.PRL.103.103001,Guthohrlein01.Nature.414.49,Steiner13.PRL.110.043003}. However, the use of cavities involves technical challenges such as [ \textit{insert} ]. In this manuscript, we demonstrate how controlled suppression of the carrier and first order sideband coupling strengths in an optical transition may be achieved without the use of a cavity. 

%By retroreflecting an incident laser field off of a surface-electrode Paul trap (Fig.~\ref{fig:schematic}), a standing wave field is formed at the location of a trapped ion. The ion's position within the standing wave fringes is controlled by the application of electric fields. This configuration utilizes a well-researched method of ion trapping without any additional components [ \textit{refs to surface trapping papers} ]. In addition, the periodicity of the coupling strengths with respect to the wavelength of the exciting field provides a measure of ion displacement as a function of applied field. We compare this measure against predictions made by a numerical model of our ion trap potentials.   

A trapped ion interacting with an ideal standing wave laser field is treated in Refs. \cite{Cirac92.PRA.46.2668,James98.APB.66.181}. We extend this treatment to include an angle of incidence $\alpha$ relative to the mirror normal (see Fig. \ref{fig:schematic}) and differing couplings of the incident and reflecting beams to the atomic transition. The latter can arise due to imperfect reflectivity of the mirror, differing polarization of the two beams, and, for quadrupole transitions such as used here, the laser beam $k$ vector dependence of the coupling. For a harmonically trapped two level atom in the rotating wave approximation, the interaction Hamiltonian (up to a global phase) is
\begin{align*}
\hat{H}_I =&\ \frac{\hbar}{2}\ \hat{\sigma}_+\,  e^{-i (\Delta t - \phi)} \\
&\times \bigg\{\Omega_1  \exp \big[-i \gamma + i \eta_1 (\hat{a} e^{i \nu t} + \hat{a}^{\dagger} e^{-i \nu t})\, \big] \\
&+\ \Omega_2 \exp \big[ i \gamma + i \eta_2 (\hat{a} e^{i \nu t} + \hat{a}^{\dagger} e^{-i \nu t})\ \big] \bigg\}\ + \text{h.c.}
\end{align*}
Here, $\Omega_1$ and $\Omega_2$ are the Rabi frequencies of the incident and reflected beams, respectively, $\hat{\sigma}_+$ ($\hat{\sigma}_-)$ is the raising (lowering) operator of the two level atom, $\hat{a}^{\dagger}$ ($\hat{a}$) is the raising (lowering) operator of the secular motion, $\Delta = \omega_{laser} - \omega_0$ is the laser's detuning from the carrier resonance, and $\nu$ is the secular frequency. The phase $\gamma = ky \cos \alpha$ represents the optical phase of the standing wave at the atom's equilbrium position $y$. The Lamb-Dicke parameters $\eta_1$ and $\eta_2$ are defined as
\begin{align*}
\eta_j =  k \sqrt{\frac{\hbar}{2 m \nu}}\ \big[ \sin \theta \sin \alpha - (-1)^j \cos \theta \cos \alpha \big]
\end{align*}
where $\theta$ is the angle of the motional mode axis relative to surface normal, $m$ is the ion mass, and $k$ is the wavenumber of the gate beam. In the Lamb-Dicke approximation, the interaction Hamiltonian can be decomposed into a carrier term
\begin{align*} 
\hat{H}_{car} =&\ \frac{\hbar}{2}\ \hat{\sigma}_+\,  e^{-i\Delta t}\, e^{i \phi} \\
&\times \bigg\{\Omega_1\, \big[1 - \frac{\eta_1^2}{2} (1 + 2\, \hat{a}^{\dagger} \hat{a})\, \big]\, e^{-i \gamma}  \\
&+\ \Omega_2\, \big[1 - \frac{\eta_2^2}{2} (1 + 2\, \hat{a}^{\dagger} \hat{a})\, \big]\, e^{i \gamma} \bigg\}\ + \text{h.c.} \text{,} 
\end{align*}
a red sideband term
\begin{align*} 
\hat{H}_{rsb} &= \frac{i \hbar}{2}\ \hat{\sigma}_+ \hat{a} \ e^{-i(\Delta + \nu) t}\, e^{i \phi} \big( \Omega_1 \, \eta_1 \, e^{-i \gamma} + \Omega_2 \, \eta_2 \, e^{i \gamma} \big)\ \\
&+\ \text{h.c.} \text{,} 
\end{align*}
and a corresponding blue sideband term. Following the treatment in Ref. \cite{Roos00.Thesis}, these Hamiltonians lead to respective coupling strengths
\begin{equation}
\label{eqn:OmegaCar}
\begin{split}
\Omega_{car} =\ &\Omega_1 \  e^{-i \gamma} \left [ 1 - \frac{\eta_1^2}{2}(1 + 2n)\, \right ] \\
+\ &\Omega_2 \  e^{i \gamma} \left [ 1 - \frac{\eta_2^2}{2}(1 + 2n)\, \right ]
\end{split}
\end{equation}
and
\begin{equation}
\label{eqn:OmegaSB}
\begin{split}
&\Omega_{rsb} = i \sqrt{n} \left( \Omega_1\, e^{-i \gamma}\, \eta_1 +\Omega_2\, e^{i \gamma}\, \eta_2 \right ) \\
&\Omega_{bsb} = i \sqrt{n+1} \left( \Omega_1\, e^{-i \gamma}\, \eta_1 +\Omega_2\, e^{i \gamma}\, \eta_2 \right )
\end{split}
\end{equation}
where $n$ is the occupation number of the quantized harmonic oscillator. Interference between the $e^{\pm i \gamma}$ terms produce fringes in the coupling strengths as the ion position changes.  When $\cos \theta \cos \alpha > \sin \theta \sin \alpha$, $\eta_1$ and $\eta_2$ have opposite sign, and the carrier and sideband fringes are $180^{\circ}$ out of phase. %For the quadrupole transition used for this demonstration, $\Omega_1$ and $\Omega_2$ due to the dependence on the laser propogation direction.  
When the atom lies on a node of the standing wave, $\gamma = l \pi$ (with $l$ an integer) and the carrier coupling strength is maximized while the sideband coupling strengths are minimized. On an anti-node, $\gamma = (l + 1/2) \pi$ and the converse is true. These fringes correspond to a physical displacement of the ion by $y = \lambda/(2 \cos \alpha)$, where $\lambda$ is the wavelength of the exciting laser.

%\begin{figure}[ht]
%    \begin{center}
%        \includegraphics{figure2}
%		\end{center}
%     \caption{\label{fig:leveldiagram} Level diagram of $^{40}$Ca$^+$. The $S_{1/2} \rightarrow D_{5/2}$ quadrupole transition at $\lambda = 729$ nm (red) is used for measuring the %effects of the standing wave field on carrier and sideband transitions. The The $S_{1/2} \rightarrow P_{1/2}$ transition ($\lambda = 397$ nm) is used for Doppler cooling, state detection, %and state preparation.  }
%\end{figure}

\begin{figure*}
    \begin{center}
        \includegraphics{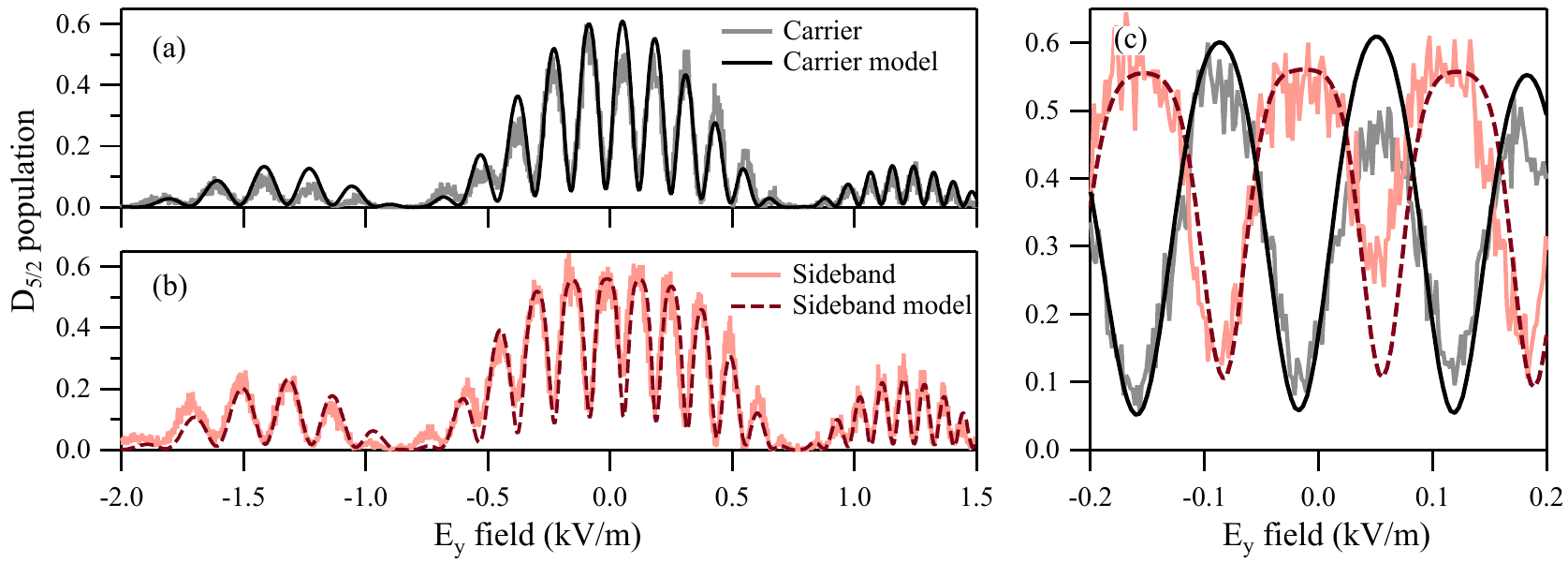}
        \caption{\label{fig:EyScan} $D_{5/2}$ populations vs. applied $E_y$ field measured after (a) carrier and (b) red sideband transitions. The sideband's gate beam power is 9.5~dB greater than the carrier's, and the interaction time is fixed for both cases (13~$\mu$s).  The overall envelope is generated by the micromotion modulation $J_0(\kappa)$ (see text). (c) The carrier and sideband populations oscillate 180$^o$ out of phase as the ion is transported through the standing wave fringes. The solid lines represent a fit to the data using the model described in the text. The deviation between the data and fit at $E_y\approx0.05$~kV$/$m may be due to scattering of the reflected light from the trap surface.}
    \end{center}
\end{figure*}

Our experimental configuration is depicted schematically in Fig. \ref{fig:schematic}(b). We use a surface electrode linear ion trap to confine and cool $^{40}$Ca$^+$ ions, as described in Ref. \cite{Doret12.NJP.14.073012}. The ion is confined $\sim$60 $\mu$m above the trap surface by a combination of RF and DC potentials.  We use the $|S_{1/2}, m_F = -\frac{1}{2} \rangle \rightarrow |D_{5/2}, m_F = -\frac{5}{2} \rangle$ quadrupole transition at $\lambda = 729$ nm for our measurements. The 729 nm beam reflects off of the aluminum trap surface with a reflectivity of $86\%$, producing a standing wave field in the $y$ direction. The angle $|\alpha|\approx20^{\circ}$ due to restrictions in the beam path, which limits the fringe contrast. We tune the laser to address either the carrier transition or the first order red sideband transition of a motional mode with $\nu=2\pi \times 4.75$ MHz (on the RF null) and $\theta = 13^{\circ}$ (see Fig. \ref{fig:schematic}(b)).

To displace the ion's equilibrium position along the standing wave, we apply an electric field $E_y$ by adjusting the trap's DC potentials. The case $E_y$ = 0 corresponds to the ion being on the RF null. For $E_y>0$ and $E_y<0$, the ion is displaced along $+y$ and $-y$, respectively. The range of $E_y$ values used in this experiment displaces the ion over a 10 $\mu$m range. The fringes in the carrier and sideband transitions resulting from this displacement are evident in Fig.~\ref{fig:EyScan}.

Displacing the ion from its equilibrium position on the RF null affects the dynamics in several ways beyond the fringing. First, the displacement introduces micromotion \cite{Leibfried03.RMP.75.281} that modulates the coupling strengths of both the carrier and sideband \cite{Berkeland98.JAP.83.5025}. To account for this modulation, we multiply equations \eqref{eqn:OmegaCar} and \eqref{eqn:OmegaSB} by the Bessel function $J_0(\kappa)$, where $\kappa$ is the modulation parameter given by
\begin{equation}
\label{eqn:kappa}
\kappa = \cos \beta \frac{2}{\lambda\, \nu_{RF}} \sqrt{ \frac{q\, \Phi_{pp}(x,y)}{m}  } .
\end{equation}
Here $\lambda$ is the wavelength of the gate beam, $\nu_{RF}$ is the trap RF frequency, $m$ and $q$ are the mass and charge of the ion, $\Phi_{pp}(x,y)$ is the RF pseudopotential, and $\beta$ is the angle between the micromotion direction and the gate beam. Our trap has an RF quadrupole that is rotated in the $xy$ plane relative to the surface normal \cite{Doret12.NJP.14.073012}; during ion displacement, we apply a small $E_x$ proportional to $E_y$ such that $\beta = 0$ at all points. Second, the ion's motional frequencies change after displacement. We track the changing frequency by measuring the sideband's resonance for different displacements.

Figs. \ref{fig:EyScan}(a) and (b) show $D_{5/2}$ state populations measured after driving carrier or sideband transitions as a function of applied $E_y$ field. The $D_{5/2}$ population is determined by the electron shelving technique that correlates observation fluorescence with the state of the ion: $S_{1/2}$ is bright and $D_{5/2}$ is dark. We observe fringes due to the standing wave field superimposed with the $J_0(\kappa)$ envelope due to the micromotion modulation. The maxima (minima) of the carrier fringes correspond to the ion positioned on nodes (anti-nodes) of the standing wave. For a dipole coupling instead of the quadrupole used here, the node/anti-node dependence is reversed. The carrier and sideband fringes are overlaid in Fig. \ref{fig:EyScan}(c). The standing wave fringes of the carrier and sideband oscillate 180$^{\circ}$ out of phase with one another, as predicted by equations \eqref{eqn:OmegaCar} and \eqref{eqn:OmegaSB}. 

\begin{figure}[tbh]
    \begin{center}
        \includegraphics{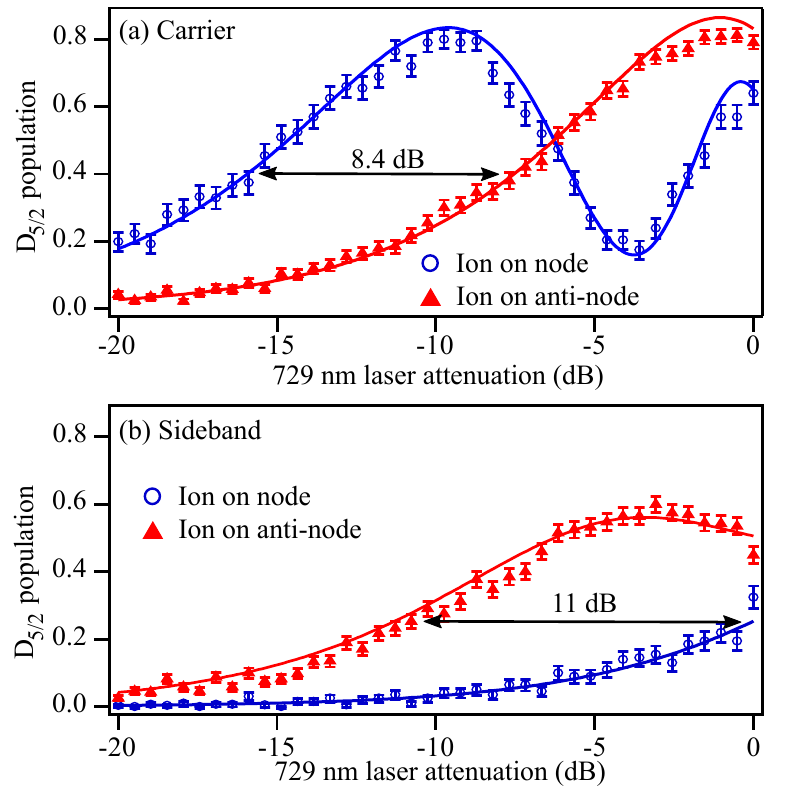}
        \caption{\label{fig:PowerScans} $D_{5/2}$ populations vs. relative gate beam power measured after (a) carrier and (b) red sideband transitions. Interaction time is fixed for both cases (13~$\mu$s). The populations were measured with the ion on either a node (blue circles, $E_y = -0.08$ kV/m) or an adjacent anti-node (red triangles, $E_y = -0.01$ kV/m) near the RF null ($J_0(\kappa) \approx 1$). An effective suppression of 8 dB for the carrier and 11 dB for the sideband is achieved. The solid lines represent a simultaneous fit to the data using the model described in the text. The results from Fig.~\ref{fig:EyScan}(a) and (b) were acquired at -13.5 dB and -4 dB, respectively.}
    \end{center}
\end{figure}

Fig. \ref{fig:PowerScans}(a) shows that by adjusting the ion's position in the standing wave laser field, we can achieve an effective suppression of the carrier that is equivalent to an 8 dB reduction in the gate beam power. For the sideband in Fig. \ref{fig:PowerScans}(b), an equivalent 11 dB reduction can be achieved. 

To produce the fits seen in Figs. \ref{fig:EyScan} and \ref{fig:PowerScans}, we use the excited state population equation for a thermal ion given by 
\begin{equation*}  
P( D_{5/2} ) = \sum\limits_{n=0}^{\infty} \frac{\bar{n}^n}{(\bar{n} + 1)^{n+1}} \sin^2 \left[ \frac{1}{4}\, \Omega(\gamma, n)\, J_0(\kappa)\, t \right]\, \text{,}
\end{equation*}
where $\bar{n}$ is the ion's mean number of motional quanta and $\Omega(\gamma, n)$ is $\Omega_{car}$ or $\Omega_{rsb}$ depending on whether we are fitting carrier or sideband data, respectively \cite{Leibfried03.RMP.75.281}. The fringes and overall envelope  in Fig. \ref{fig:EyScan} arise from the dependence of $\gamma$ and $\kappa$ on the ion's displacement $y$. We parameterize $\gamma = k y \cos \alpha$ with $y = \sum_{j=0}^4 a_j E_y^j$  and parameterize $\kappa^2 = \sum_{j=2}^4 m_j E_y^j\,$, for coefficients $a_j$ and $m_j$. These coefficients,  along with $\Omega_{1}$, $\Omega_{2}$, $\alpha$, and $\bar{n}$, form the full set of parameters that define the simultaneous fits of the data in Figs.~\ref{fig:EyScan} and \ref{fig:PowerScans}. In particular, the set has $|\alpha|=18^{\circ}$, $\Omega_2\, / \, \Omega_1=0.52$, and $\bar{n}=18$ (consistent with the Doppler cooled ion in our configuration).

\begin{figure}[tb]
    \begin{center}
        \includegraphics{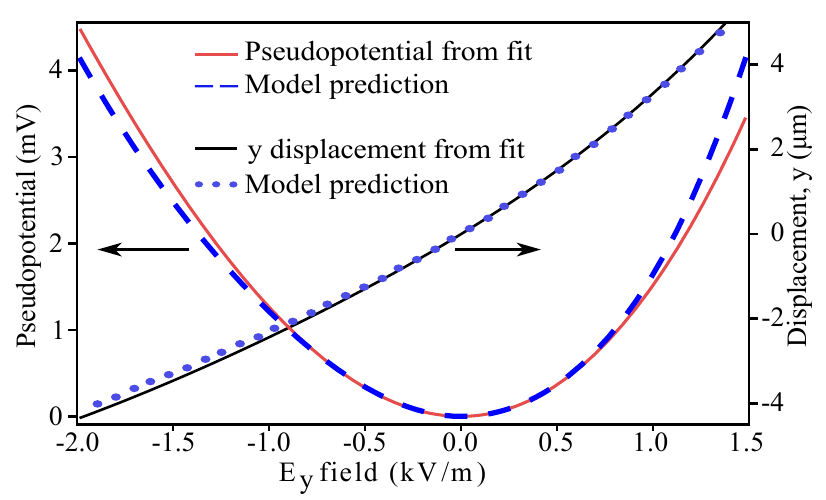}
        \caption{\label{fig:model} Dependence of the pseudopotential $\Phi_{pp}$ and ion displacement $y$ on the applied field $E_y$, along with predictions from a numerical model of the trapping potentials. Deviations from the model can be attributed non-uniformity of the stray fields in the trapping region. }
    \end{center}
\end{figure}

Using equation \eqref{eqn:kappa}, we can determine the pseudopotential $\Phi_{pp}$ from $\kappa^2$.  Fig. \ref{fig:model} plots the resulting $\Phi_{pp}$ and the ion displacement $y$ as a function of $E_y$. We compare these with the results from a numerical model of the trapping potentials. The model includes a 700 V/m uniform stray field that was adjusted to match the observed RF null in Fig.~\ref{fig:EyScan}. The magnitude of the RF trapping potential used in the model was adjusted to match the observed mode frequencies at $E_y=0$. There are no other free parameters used in the model. The agreement seen provides a measure of confidence in the model and future predictions. 

To summarize: we have demonstrated how the carrier and motional sideband transition coupling strengths of a trapped ion may be controlled by displacing the ion within a resonant standing wave field. In our configuration, in which a standing wave is formed by retroreflecting an incident beam off of a surface electrode trap, we achieve suppression of the carrier (sideband) by the equivalent of an 8 dB (11 dB) reduction in the driving beam power. Our experimental results are in good agreement with both theoretical models of the coupling strengths' behaviors and numerical models of the trapping potentials. The degree of carrier/sideband suppression achievable ultimately depends on the quality of the standing wave field at the ion's position, which is related to the ratio $\Omega_2\, / \, \Omega_1$. The trap's surface reflectivity limits us here to  $\Omega_2\, / \, \Omega_1=\sqrt{0.86}=0.93$. We can approach this limit with improved beam alignment, which would allow for a carrier suppression equivalent to a 29 dB reduction in driving beam power when $\bar{n}\approx0$ and $|\alpha|<10^{\circ}$. Further carrier suppression would require the use of a more reflective surface. For instance, we measure the reflectivity from a gold coated trap like the one described Ref. \cite{Guise15.JAP.117.174901} to be $>98\%$, which could provide an equivalent carrier suppression of $>\,$40 dB. Similar quality standing waves could be generated by incorporating a metallic mirror adjacent to an ion trap if the trap surface is not amenable to this purpose. 

Suppression of the carrier implies that the driving laser's power may be freely increased by an amount equivalent to the effective suppression, allowing for faster sideband interactions with no increased chance of an off-resonant carrier excitation. For the 8 dB effective carrier suppression reported here, sideband interactions could be performed 2.5 times faster; at the 29 dB suppression limit of our aluminum trap, 28 times faster; at the $>\,$40 dB suppression limit of a gold coated trap, $>\,$100 times faster. 

In our current configuration, a 29 dB carrier suppression factor would reach the regime in which simulating the expansion of the universe with trapped ions becomes experimentally feasible, such that the excitation of detectors occurs when cosmic photons are created \cite{Menicucci10.NJP.12.095019}. When the beam has more running wave character, the simulation is dominated by excitation of detectors with photon creation or destruction. Quantum simulations of the expanding universe have also been proposed using Bose-Einstein condensates \cite{PhysRevA.70.063615} and can be performed in a digital manner with a number of quantum systems \cite{RevModPhys.86.153,Mezzacapo2014}.

% Acknowledgements

\begin{acknowledgments}
This work was supported by the Georgia Tech Research Institute. KRB was supported by the Director of National Intelligence (ODNI), Intelligence Advanced Research Projects Activity (IARPA), under US Army Research Office (ARO) contracts W911NF-10-1-0231. TD would like to thank support by a Georgia Tech President's Fellowship.
\end{acknowledgments}

%\bibliography{paper}
% Bibliography 
\bibliography{paper}

\end{document}